\documentclass[aps,prd,amssymb,amsmath,nofootinbib,superscriptaddress]{revtex4}

\usepackage{graphicx}

\begin{document}

\title{Gravitational wave detectors based on matter wave
interferometers (MIGO) \linebreak[3]
are no better than laser interferometers (LIGO)}

\author{Albert Roura}
\author{Dieter R. Brill}
\author{B.~L. Hu}
\author{Charles W. Misner}
\affiliation{Department of Physics, University of Maryland,
College Park, Maryland 20742-4111}
\author{William D. Phillips}
\affiliation{Department of Physics, University of Maryland,
College Park, Maryland 20742-4111}
\affiliation{National Institute of Standards and Technology,
Gaithersburg, MD 20899-8424}

\begin{abstract}
We show that a recent claim that matter wave interferometers have a
much higher sensitivity than laser interferometers for a comparable
physical setup is unfounded. We point out where the mistake in the
earlier analysis is made. We also disprove the claim that only a
description based on the geodesic deviation equation can produce the
correct physical result. The equations for the quantum dynamics of
non-relativistic massive particles in a linearly perturbed spacetime
derived here are useful for treating a wider class of related physical
problems. A general discussion on the use of atom interferometers for
the detection of gravitational waves is also provided.
\end{abstract}


\maketitle

\section{Introduction}
\label{sec1}

Chiao and Speliotopoulos proposed in a recent paper \cite{chiao04a}
that gravitational wave detectors based on matter wave interferometry
(MIGO) have sensitivity, under equal conditions, far superior to those
based on laser interferometers such as LIGO. They argue that this is
true especially for high-frequency gravitational waves and that it is
better to use slow rather than fast atoms (see also
Ref.~\cite{foffa04a} for related work).  In a separate paper
\cite{speliotopoulos04a} Chiao and Speliotopoulos also argued that the
usual quantum mechanical description for non-relativistic particles
moving in the spacetime of a linearized gravitational wave
\cite{linet76,stodolsky79,cai89,borde01,borde04} is inconsistent.
They suggest that the description in terms of geodesic coordinates
(defined precisely in the next section) and that based on the geodesic
deviation equation in the rigid frame are inequivalent. In this paper,
we show that all those claims are incorrect. We also take this
opportunity to address some general features of matter wave
interferometry in comparison to light interferometry as applied to
gravitational wave detection considerations.

In the first part of this paper we show that the quantum mechanical
description in terms of geodesic coordinates and that based on the
geodesic deviation equation in the rigid frame are equivalent, contrary
to Chiao and Speliotopoulos's claim. They just correspond to the
use of two different coordinate systems and are related by a
time-dependent unitary transformation. In the second part of this
paper we compute the phase shift for an atom interferometer due to the
passage of a gravitational wave in each of these two coordinate
systems and show explicitly that the two results are the same (as one
could have implicitly inferred from the equivalence of the two
descriptions). This contradicts the results of Chiao and
Speliotopoulos in Ref.~\cite{speliotopoulos04b}, where different
expressions were obtained explicitly for the two approaches. We point
out that their mistake originated from their procedure for computing
the phase shift, resulting in different answers for the two coordinate
systems.

To help readers identify the source of confusion (perhaps at some risk
of oversimplifying), we can explain their mistake as follows.  Chiao
and Speliotopoulos obtain the phase difference between the two ends of
an interferometer arm by computing the total action for the classical
trajectory with a given initial velocity that goes from one end of the
arm to the other. In fact, they are interested in the change of the
action due to the interaction with the gravitational wave, which is
treated perturbatively. They keep the initial and final times
fixed. However, when the perturbation is included, the position at the
final time no longer coincides with the second end of the arm since
the initial velocity and position also remain fixed. The correct
result is obtained when the change of the initial time needed for
preserving the boundary conditions is taken into account.

When the correct analysis is employed, the great enhancement in the
sensitivity of MIGO compared to LIGO found by Chiao and Speliotopoulos
no longer exists, and it becomes clear that the sensitivity of MIGO is
higher for faster atoms. Furthermore, despite the fact that the two
cases correspond to slightly different physical situations (the atoms
in MIGO are non-relativistic, whereas photons are highly
relativistic), the expression for the phase shift in an atom
interferometer that we obtain turns out to be formally equivalent to
the general expression (for arbitrary gravitational wave frequencies)
for a laser interferometer (if one replaces the laser wavelength by
the de Broglie wavelength\footnote{Throughout this paper by \emph{de
Broglie wavelength} of an atom interferometer we mean the
wavelength that free atoms with energy $E$ would have in the absence
of gravitational waves, where $E$ is the constant energy at which the
atoms are emitted by the interferometer source.} $\lambda$ and $c$ with
$v$) except for an additional term that accounts for the fact that the
arms of MIGO are rigid, in contrast to laser interferometers such as
LIGO, whose mirrors are freely suspended. (The implications of having
a MIGO with freely suspended mirrors are discussed in
Sec.~\ref{sec5}).

Whereas LIGO operates in the low frequency regime (transit time
smaller than the gravitational wave period), for most relevant
situations MIGO would operate in the high frequency regime. In that
regime, the term in the expression for the phase shift due to having
rigid arms dominates.  That term is given by $L/\lambda$, where $L$
is the length of the arm, times an oscillatory factor of order one which
is a harmonic function of the time times $\omega$ for a gravitational
wave with angular frequency $\omega$. Therefore, under equal
conditions, \emph{i.e.}, the same values for the wavelength, the
length of the interferometer arms and the flux, the sensitivity of
MIGO would be comparable to that of LIGO.

Going beyond the correction of mistakes in previous analysis on the
same problem, one can still ask whether there is a feasible range of
parameters (wavelength, length of the interferometer arms, and flux)
for an atom interferometer such that its sensitivity for the detection
of gravitational waves is better than that of LIGO. We will briefly
discuss this point in Sec.~\ref{sec5}. With these discussions as a
start, it is our hope that some interest in both the gravitational
physics and the AMO (atomic, molecular and optical physics) communities
will be generated to explore in broader terms the scientific potential
of atom interferometry toward precision measurements related to
gravity effects.

The paper is organized as follows. In Sec.~\ref{sec2} we derive the
geodesic equations describing the motion of free-falling particles in
two different coordinate systems (the geodesic versus the rigid), and
show how they are related to each other by a time-dependent local
transformation. In Sec.~\ref{sec3} we write down the action for the
non-relativistic motion of a massive particle in the spacetime of a
linearized gravitational wave and derive the Hamiltonian in terms of
the canonical variables for these two coordinate systems. In the
quantum theory obtained from canonical quantization, we write down the
Schr\"odinger equations for the two coordinate systems used and show
that they produce unitarily related dynamics. This disproves the claim
that only the rigid coordinates give the correct physical results. In
Sec.~\ref{sec4} we derive the phase difference between the ends of an
interferometer arm including the effect due to gravitational waves
from an analysis of the Hamilton-Jacobi equation, which corresponds to
the lowest order approximation for the Schr\"odinger equation in a
JWKB semiclassical expansion, with the action expanded in powers of
the metric perturbation. We show explicitly that the results for the
phase differences calculated in these two coordinate systems coincide
and identify where the error was made in earlier works.  In
Sec.~\ref{sec5} we discuss the physical implications of the result
obtained in Sec.~\ref{sec4} as well as some general aspects about the
use of atom interferometers for the detection of gravitational waves.
Some technical details about an alternative way to compute the phase
differences are included in Appendix~\ref{appA}. They are useful in
order to compare with Chiao and Speliotopoulos's approach. Finally, in
Appendix~\ref{appB} we explain why, contrary to the claim made in
Ref.~\cite{foffa04b}, the effects due to the reflection off the
mirrors cancel out in the final result for the phase shift.

Throughout this paper relativistic units with $c=1$ are used. Indices
are raised and lowered using the background Minkowski metric
$\eta_{\mu\nu}$. Greek indices correspond to spacetime indices running
from $0$ to $3$, whereas Latin indices correspond to spatial indices
running from $1$ to $3$. The Einstein summation convention over
repeated indices is employed.

\section{Geodesic versus rigid coordinates for a linearized
gravitational wave}
\label{sec2}

In the transverse-traceless gauge, the metric for a linearized
gravitational wave propagating along the $z$ direction is given by
\begin{equation}
ds^2 = -dt^2 + (\delta_{ij} + h_{ij} (t - z)) dX^idX^j
\label{metric1},
\end{equation}
where $X^i$ correspond to three spatial coordinates (in particular $z
\equiv X^3$) and $h_{ij}$ is a tensor-valued function with arbitrary
dependence on $t - z$ that satisfies the conditions $h_{i3} = 0$ and
$h^i_i = 0$ \cite{misner73}.

The geodesic equations describing the motion of free-falling particles
in this coordinate system are
\begin{eqnarray}
\frac{d^2 X^i}{d\tau^2} &=& - \dot{h}^i_j \frac{dt}{d\tau}
\frac{d X^j}{d\tau} + \dot{h}^i_j \frac{dz}{d\tau} \frac{d X^j}{d\tau}
- \frac{1}{2} \delta^i_3\, \dot{h}_{jk} \frac{d X^j}{d\tau}
\frac{d X^k}{d\tau} + O(h^2_{ij}) \label{geodesic1}, \\
\frac{d^2 t}{d\tau^2} &=& -\frac{1}{2} \dot{h}_{ij} \frac{d X^i}{d\tau}
\frac{d X^j}{d\tau} + O(h^2_{ij}) \label{geodesic2},
\end{eqnarray}
where $\tau$ is the proper time and the overdot denotes a derivative
with respect to the inertial time $t$. The initial conditions must
satisfy the normalization condition $(dt/d\tau)^2 - (\delta_{ij} +
h_{ij}) (dX^i/d\tau)(dX^j/d\tau) = 1$, which is then preserved up to
linear order in $h_{ij}$ by Eqs.~(\ref{geodesic1})-(\ref{geodesic2}).
For a non-relativistic particle we have $(dt/d\tau) = 1 + O(v^2)$ with
$v^2 = (dX^i/dt)(dX_i/dt)$, and Eq.~(\ref{geodesic1}) becomes
\begin{equation}
\frac{d^2 X^i}{dt^2} = - \dot{h}^i_j \frac{d X^j}{dt}
+ O(h^2_{ij},v^2h_{ij}) \label{geodesic3}.
\end{equation}
Since worldlines with $X^i (t) = \mathrm{constant}$ are solutions of
the non-relativistic equation~(\ref{geodesic3}), or even
Eq.~(\ref{geodesic1}), we will use the term \emph{geodesic
coordinates} for this coordinate system (commonly referred as
\emph{transverse-traceless coordinates}).

Alternatively, one can work with a new coordinate system related to
$(t, X^i)$ by a transformation
\begin{eqnarray}
t &=& t , \\
x^i &=& X^i + \frac{1}{2} h^i_j X^j
\label{change},
\end{eqnarray}
which is generated by the vector field $\xi^{\mu}(t,X^i) = - (1/2)
\eta^{\mu i} h_{ij}(t - z) X^j$. The metric perturbation in this new
coordinate system becomes $\tilde{h}_{\mu\nu} = h_{\mu\nu} +
\partial_\mu \xi_\nu + \partial_\nu \xi_\mu$ and hence the full
metric is given by
\begin{equation}
ds^2 = -dt^2 - \dot{h}_{ij}(t - z) x^{i} dt dx^j + \delta_{ij} dx^idx^j
+ \dot{h}_{ij}(t - z) x^{i} dz dx^j + O(h_{ij}^2)
\label{metric2},
\end{equation}
where terms of quadratic or higher order in $h_{ij}$ have been
neglected. It should be noted that this coordinate system is only
consistent with the linearized treatment of the metric perturbation as
long as $\dot{h}_{ij}(t) x^{i}$ is of the same order as
$h_{ij}$. Therefore, if we consider a gravitational wave of wavelength
$\lambda_\mathrm{GW}$, we should restrict our consideration to a
region such that $x^i \lesssim \lambda_\mathrm{GW}$ for any $i$.

The geodesic equations in the $(t, x^i)$ coordinates are
\begin{eqnarray}
\frac{d^2 x^i}{d\tau^2} &=& \frac{1}{2} \ddot{h}^i_j x^j
\left( \frac{dt}{d\tau} \right)^2
- \frac{1}{2} \ddot{h}^i_j x^j \frac{dt}{d\tau} \frac{dz}{d\tau}
+ \frac{1}{2} \ddot{h}^i_j x^j \left( \frac{dz}{d\tau} \right)^2
- \frac{1}{2} \delta^i_3\, \dot{h}_{ij} \frac{d x^i}{d\tau}
\frac{d x^j}{d\tau} + O(h^2_{ij}) \label{geodesic4}, \\
\frac{d^2 t}{d\tau^2} &=& - \frac{1}{2} \dot{h}_{ij} \frac{d x^i}{d\tau}
\frac{d x^j}{d\tau} + O(h^2_{ij})
\label{geodesic5}.
\end{eqnarray}
Proceeding analogously to the previous case, one obtains the following
equation for a non-relativistic particle:
\begin{equation}
\frac{d^2 x^i}{dt^2} = \frac{1}{2} \ddot{h}^i_j x^j
- \frac{1}{2} \ddot{h}^i_j x^j \frac{dz}{d\tau}
+ O(h^2_{ij},v^2h_{ij}) \label{geodesic6}.
\end{equation}
{F}rom Eqs.~(\ref{geodesic6}) or (\ref{geodesic3}) one can see that in
the non-relativistic regime it is possible to have solutions of the
geodesic equations in both coordinate systems with $z(t) =
\mathrm{constant}$. For simplicity, from now on we will restrict
ourselves to the case in which the motion of the non-relativistic
particle is confined to the $z = 0$ plane. Eq.~(\ref{geodesic6})
reduces then to
\begin{equation}
\frac{d^2 x^i}{dt^2} = \frac{1}{2} \ddot{h}^i_j x^j
+ O(h^2_{ij},v^2h_{ij}) \label{geodesic7}.
\end{equation}
Since one can verify that the physical distance between worldlines
with $x^i (t) = \mathrm{constant}$ remains constant in time up to
quadratic order in $h_{ij}$, we will use the term \emph{rigid
coordinates} for this coordinate system.

Eq.~(\ref{geodesic7}) can also be derived from the geodesic deviation
equation \cite{misner73}, which governs the separation of nearby
geodesics and reduces for rigid coordinates to
\begin{equation}
\frac{d^2 x^i}{dt^2} = R^i_{\ 00j} x^j
\label{deviation},
\end{equation}
by taking into account that the Riemann tensor to linear order in the
metric perturbations is given simply by
\begin{equation}
R^i_{\ 00j} = \frac{1}{2} \ddot{h}^i_j
\label{riemann} \, .
\end{equation}
However, this is not a peculiar feature due to using rigid coordinates:
Eq.~(\ref{geodesic3}), for instance, can also be derived from the
geodesic deviation equation expressed in terms of geodesic
coordinates.

\section{Description of a non-relativistic particle propagating on a
linearized gravitational wave}
\label{sec3}

The action, from which the geodesic equations can be derived, for a
particle of mass $m$ propagating on a spacetime characterized by a
metric $g_{\mu\nu}$ is given by
\begin{equation}
S[y^\mu (\gamma)] = - m \int d\tau = - m \int d\gamma
\sqrt{- g_{\alpha\beta}(y^\mu (\gamma)) \frac{dy^\alpha}{d\gamma}
\frac{dy^\beta}{d\gamma}}
\label{action1},
\end{equation}
where $\gamma$ is an arbitrary parameter for the particle's worldline.
Here we used the notation $y^\mu$ for the spacetime coordinates in a
general coordinate system to distinguish them from the particular
coordinate systems used to describe a linearized gravitational wave
introduced in the previous section. The action for a non-relativistic
particle propagating on a linearly perturbed Minkowski spacetime with
metric $g_{\mu\nu}(x) = \eta_{\mu\nu} + \tilde{h}_{\mu\nu}(y)$, can be
derived by taking the particular parametrization $\gamma = y^0 \equiv
t$ and neglecting terms of order $\tilde{h}^2_{ij}$, $v^3$ or
higher. In particular, if $\tilde{h}_{00} = 0$ (as will be the case in
the situations addressed in this paper) the non-relativistic action
becomes
\begin{equation}
S[y^i (t)] = \int dt L \left( y^i(t), \dot{y}^i(t) \right)
= \int dt \, \frac{m}{2} \left( \frac{dy^i}{dt} \frac{dy_i}{dt}
+ 2 \tilde{h}_{0i}(t,y^i) \frac{dy^i}{dt}
+ \tilde{h}_{ij}(t,y^i) \frac{dy^i}{dt} \frac{dy^j}{dt} - 2 \right)
\label{action2}.
\end{equation}
The last term in the integrand can be rewritten as a time derivative of
$m \, t$ and does not affect the equations of motion derived from the
action.

We now consider the spacetime of a linearized gravitational wave in
the two coordinate systems introduced in the previous section.
Recall that we restrict our attention to the case in which the motion
is confined to the $z = 0$ plane.

\subsection{Geodesic coordinates}
\label{sec3.1}

When geodesic coordinates are used, the metric is given by
Eq.~(\ref{metric1}) and the Lagrangian in Eq.~(\ref{action2}) reduces
to
\begin{equation}
L(X^i,\dot{X}^i) = \frac{m}{2} \left( \dot{X}^i \dot{X}_i
+ h_{ij}(t) \dot{X}^i \dot{X}^j - 2 \right) \label{lagrangian1}.
\end{equation}
The equation of motion derived from this Lagrangian coincides, as it
should, with Eq.~(\ref{geodesic3}), which was obtained directly from
the geodesic equation.

One can easily make the transition to the canonical formalism. With
the conjugate momenta $P_i = m (\dot{X}_i + h_{ij}(t) \dot{X}^j)$ one
obtains the following result for the Hamiltonian up to linear order in
the metric perturbation:
\begin{equation}
H_\mathrm{g}(X^i,P_i) = \frac{1}{2m} P_i P^i
- \frac{1}{2m} h_{ij}(t) P^i P^j + m
\label{hamiltonian1}.
\end{equation}

\subsection{Rigid coordinates}
\label{sec3.2}

Alternatively, if the rigid coordinates are employed, the metric is
given by Eq.~(\ref{metric2}), and the Lagrangian for the motion of a
non-relativistic particle becomes
\begin{equation}
L(x^i,\dot{x}^i) = \frac{m}{2} \left( \dot{x}^i \dot{x}_i
- \dot{h}_{ij}(t) x^j \dot{x}^i - 2 \right)
\label{lagrangian2}.
\end{equation}
The equation of motion derived from this Lagrangian coincides
with that obtained from the geodesic equation and given by
Eq.~(\ref{geodesic7}).

Again, one can easily make the transition to the canonical formalism.
The conjugate momenta are $p_i = m (\dot{x}_i - (1/2) \dot{h}_{ij}(t) x^j)$
and the Hamiltonian up to linear order in the metric perturbation is
\begin{equation}
H_\mathrm{r}(x^i,p_i) = \frac{1}{2m} p_i p^i
+ \frac{1}{2} \dot{h}_{ij}(t) x^i p^j + m
\label{hamiltonian2}.
\end{equation}

It should be noted that the last term in Eqs.~(\ref{hamiltonian1}) and
(\ref{hamiltonian2}) simply corresponds to the rest energy of the
massive particle and can be absorbed by redefining the origin of
energies. When doing so, at the quantum level one should also redefine
the state vector $|\psi (t)\rangle$ to $\exp (i m t) |\psi
(t)\rangle$. From now on we will presume that such redefinitions have
been made.

\subsection{Equivalence between the two descriptions}
\label{sec3.3}

The two Lagrangians are equivalent, up to linear order in the metric
perturbation, under the time dependent coordinate transformation
\begin{equation}
X^i \rightarrow x^i = X^i + \frac{1}{2} h^i_j(t) X^j
\label{change2}.
\end{equation}
This change of
coordinates induces the following transformation for the momenta up to
linear order in the metric perturbation:
\begin{equation}
P_i \rightarrow p_i = P_i - \frac{1}{2} h_{ij}(t) P^j
\label{change3},
\end{equation}
which together with Eq.~(\ref{change2}) constitutes a time
dependent canonical transformation, \emph{i.e.}, it leaves the
symplectic structure of phase space (or, equivalently, the Poisson
brackets for the position and momentum variables) invariant at every
instant of time, up to that order. Under this canonical transformation
($X^i \rightarrow x^i,\ P_i \rightarrow p_i$) the Hamiltonian
transforms, up to linear order in $h_{ij}$, as
\begin{equation}
H_\mathrm{g}(X^i,P_i) \rightarrow H_\mathrm{r}(x^i,p_i)
= H_\mathrm{g}(X^i,P_i) + \frac{1}{2} \dot{h}_{ij}(t) X^i P^j 
\label{change4},
\end{equation}
where the subscripts $\mathrm{g}$ and $\mathrm{r}$ denote quantities
in geodesic and rigid coordinates respectively, but
$P_i\dot{X}^i(X^j,P_j) - H(X^i,P_i)$ remains invariant.

When quantizing the theory by promoting the position and momentum
variables to operators satisfying the canonical commutation relations,
the previous canonical transformation corresponds to an analogous
transformation for the position and momentum operators that preserves
their commutation relations. According to the Stone-Von Neumann
theorem (see, for instance, Ref.~\cite{wald94} and references
therein), this is sufficient to guarantee that the two representations
are unitarily equivalent, \emph{i.e.}, connected by a unitary
transformation, which will be time dependent in this case.

In fact, the unitary equivalence of the descriptions can be shown
explicitly.  Let us start with the Schr\"odinger equation in terms of
the geodesic coordinates
\begin{equation}
i \hbar \frac{d}{dt} \, _\mathrm{g} \langle y^i |\psi (t) \rangle
= {}_\mathrm{g} \langle y^i | \hat{H_\mathrm{g}} (\hat{X}^i,\hat{P}_i)
| \psi (t) \rangle
\label{schrodinger1} .
\end{equation}
where we have introduced the position basis\footnote{A mathematically
more rigorous treatment should employ strictly normalizable states,
but identical conclusions would be obtained if a basis of normalizable
wave-packets were used. In addition, we should keep in mind that, when
using rigid coordinates, the perturbative expansion in $h_{ij}$ is
only valid for regions smaller than the wavelength of the
gravitational wave, as mentioned in Sec.~\ref{sec2}. Therefore,
throughout this subsection it is implicitly assumed that the Hilbert
space is restricted to states localized within a region smaller than
$\lambda_\mathrm{GW}$.} $| y^i \rangle_\mathrm{g}$, characterized
by $\hat{X^j} | y^i \rangle_\mathrm{g} = y^j | y^i \rangle_\mathrm{g}$
and $_\mathrm{g} \langle y^i | y^{\prime i} \rangle _\mathrm{g} =
\delta^{(3)} (y^i - y^{\prime i})$. The action of the momentum
operator in this basis is given by $_\mathrm{g} \langle y^i |
\hat{P}^j | y^{\prime i} \rangle_\mathrm{g} = - i \hbar ( \partial /
\partial y^j ) \delta^{(3)} (y^i - y^{\prime i})$.  Similarly, one can
consider the position basis in rigid coordinates $| y^i
\rangle_\mathrm{r}$, which is characterized by $\hat{x}^j | y^i
\rangle_\mathrm{r} = y^j | y^i \rangle_\mathrm{r}$ and $_\mathrm{r}
\langle y^i | y^{\prime i} \rangle_\mathrm{r} = \delta^{(3)} (y^i -
y^{\prime i})$.  Taking into account that $\hat{x}^i = \hat{X}^i +
\frac{1}{2} h^i_j(t) \hat{X}^j$, one finds that the position basis in
the two descriptions are related up to linear order in $h_{ij}$ as
follows:
\begin{equation}
| y^i \rangle_\mathrm{r} = | y^i - (1/2) h^i_j(t) y^j \rangle_\mathrm{g}
\equiv \hat{U}(t) | y^i \rangle_\mathrm{g}
\label{change5},
\end{equation}
where the linear operator $\hat{U}(t)$ relating the two bases is
unitary up to linear order in $h_{ij}$ since it preserves both
orthogonality and normalization up to that order, as can be seen from
Eq.~(\ref{change5}) together with the fact that the value of the
Jacobian for the coordinate transformation in Eq.~(\ref{change2}) is
one plus terms quadratic in $h_{ij}$, which is a consequence of
$h_{ij}$ being traceless. The same unitary operator also relates the
position and momentum operators within the two descriptions in the
following way: $\hat{x}^i = \hat{U}(t) \hat{X}^i \hat{U}^{-1} (t)$ and
$\hat{p}_i = \hat{U}(t) \hat{P}_i \hat{U}^{-1} (t)$. One can
check that it is given by $\hat{U}(t) = 1 + (i/2) h_{ij} (t) \hat{X}^i
\hat{P}^j + O(h^2_{ij})$. It should be noted that the ordering of the
operator product $\hat{X}^i \hat{P}_j$ appearing in both $\hat{U}(t)$
and $\hat{H}_r$ is irrelevant because it is always contracted with
$h_{ij}$ and $h_{ij} (t) [\hat{X}^i, \hat{P}^j]$ vanishes because
$h_{ij}$ is traceless.

Furthermore, the Schr\"odinger equation in terms of rigid coordinates,
\begin{equation}
i \hbar \frac{d}{dt} \, _\mathrm{r} \langle y^i |\psi (t) \rangle
= {}_\mathrm{r} \langle y^i | \hat{H_\mathrm{r}} (\hat{x}^i,\hat{p}_i)
| \psi (t) \rangle
\label{schrodinger2} .
\end{equation}
is equivalent, under the time-dependent unitary transformation
generated by $\hat{U}(t)$, to the Schr\"odinger equation in terms of
geodesic coordinates, given by Eq.~(\ref{schrodinger1}). This can be
shown as follows. First, we rewrite $_\mathrm{r} \langle y^i |\psi
(t) \rangle$ as $\int d^3y' \, _\mathrm{r} \langle y^i | y^{\prime i}
\rangle_\mathrm{g} \, _\mathrm{g} \langle y^{\prime i} | \psi (t)
\rangle$ on the left-hand side of Eq.~(\ref{schrodinger2}), which
becomes
\begin{equation}
i \hbar \int d^3y' \left( \frac{d}{dt} \, _\mathrm{r} \langle y^i |
y^{\prime i} \rangle_\mathrm{g} \right) \, _\mathrm{g} \langle y^{\prime i}
|\psi (t) \rangle
+ i \hbar \int d^3y' \, _\mathrm{r} \langle y^i | y^{\prime i}
\rangle_\mathrm{g} \, \frac{d}{dt} \, _\mathrm{g} \langle y^{\prime i}
|\psi (t) \rangle
\label{lhs} .
\end{equation}
Next, we use the operator version of Eq.~(\ref{change4}) to substitute
$\hat{H_\mathrm{r}} (\hat{x}^i,\hat{p}_i)$ into the right-hand side of
Eq.~(\ref{schrodinger2}), which becomes
\begin{equation}
\int d^3y' \, _\mathrm{r} \langle y^i | y^{\prime i} \rangle_\mathrm{g}
\, _\mathrm{g} \langle y^{\prime i} | \hat{H_\mathrm{g}} (\hat{X}^i,\hat{P}_i)
| \psi (t) \rangle
+ \frac{1}{2} \dot{h}_{jk}(t)  \int d^3y' \,
_\mathrm{r} \langle y^i | y^{\prime i} \rangle_\mathrm{g}
\, _\mathrm{g} \langle y^{\prime i} | \hat{X}^j \hat{P}^k | \psi (t) \rangle
\label{rhs} .
\end{equation}
The last term in this expression is exactly canceled by the first term
in expression~(\ref{lhs}) since $i \hbar (d/dt) \, _\mathrm{r} \langle y^i
| y^{\prime i} \rangle_\mathrm{g} = (i \hbar / 2) \dot{h}_{jk}(t)
(\partial / \partial y') \delta (y - (1/2) \dot{h} (t) y - y') = (1/2)
\dot{h}_{jk}(t) _\mathrm{g} \langle y^{\prime i} | \hat{X}^j \hat{P}^k
| \psi (t) \rangle$, which follows from Eq.~(\ref{change5}). The whole
equation becomes then
\begin{equation}
i \hbar \int d^3y' \, _\mathrm{r} \langle y^i | y^{\prime i}
\rangle_\mathrm{g} \, \frac{d}{dt} \, _\mathrm{g} \langle y^{\prime i}
|\psi (t) \rangle
= \int d^3y' \, _\mathrm{r} \langle y^i | y^{\prime i} \rangle_\mathrm{g}
\, _\mathrm{g} \langle y^{\prime i} | \hat{H_\mathrm{g}} (\hat{X}^i,\hat{P}_i)
| \psi (t) \rangle
\label{schrodinger3} .
\end{equation}
Finally, multiplying both members by $_\mathrm{g} \langle y^{\prime
\prime i} | y^i \rangle_\mathrm{r}$ and integrating over $y^i$ one
recovers Eq.~(\ref{schrodinger1}). Therefore, the time-dependent
unitary transformation generated by $\hat{U}(t)$ accounts for the
different time evolution generated by the corresponding Schr\"odinger
equations for the two different Hamiltonians $\hat{H_\mathrm{g}}
(\hat{X}^i,\hat{P}_i)$ and $\hat{H_\mathrm{r}} (\hat{x}^i,\hat{p}_i)$.

To summarize, it is clear that, contrary to the claims made in
Refs.~\cite{speliotopoulos95,speliotopoulos04a}, the quantum mechanical
description in terms of geodesic coordinates is equivalent to that
based on the geodesic deviation equation, which simply corresponds to
using rigid coordinates, and there is no inconsistency with the usual
approach, which employs geodesic coordinates.

\section{Phase shift in atom interferometers caused by
gravitational waves}
\label{sec4}

In this section we will analyze the horizontal MIGO configuration
proposed in Ref.~\cite{chiao04a}, in which the Earth's gravitational
filed plays no major role and its effect is neglected. It is slightly
simpler than the vertical configuration also proposed there, but it
already contains all the issues that need to be addressed. Throughout
this section we will consider interferometers with rigid arms, as
proposed in Ref.~\cite{chiao04a}. Additional discussion on the
implications of having an atom interferometer with freely suspended
mirrors is provided in Sec.~\ref{sec5}.

\subsection{General expression for non-relativistic atoms}
\label{sec4.1}

The phase shift due to the passage of a linearized gravitational wave
through an atom interferometer can be computed using the geometric
optics approximation (the lowest order JWKB approximation) as long as
the wavelength of the gravitational wave and the arms of the
interferometer are much longer than the de Broglie wavelength of the
atoms. Under this approximation one only needs to solve the
Hamilton-Jacobi equation
\begin{equation}
0 = \frac{\partial S}{\partial t}
+ H \left( \frac{\partial S}{\partial y^i},y^i,h_{ij}(t) \right)
\label{HJ1},
\end{equation}
perturbatively with respect to the metric perturbation. Here we have
used a generic notation $y^i$, which can correspond to either
$X^i$ or $x^i$. Expanding the action in powers of the metric
perturbation as $S (y^i,t) = S_0 (y^i,t) + S_1 (y^i,t;h_{ij}] +
O(h^2_{ij})$, we obtain the following set of equations from
Eq.~(\ref{HJ1}):
\begin{eqnarray}
0 &=& \frac{\partial S_0}{\partial t}
+ \frac{1}{2m} \left( \frac{\partial S_0}{\partial y^i} \right)
\left( \frac{\partial S_0}{\partial y^j} \right) \delta^{ij}
\label{HJ2}, \\
0 &=& \frac{\partial S_1}{\partial t}
+ \frac{1}{m} \left( \frac{\partial S_0}{\partial y^i} \right)
\left( \frac{\partial S_1}{\partial y^j} \right) \delta^{ij}
+ H^{(1)} \left( \frac{\partial S_0}{\partial y^i} ,y^i,h_{ij}(t)
\right)
\label{HJ3},
\end{eqnarray}
where $H^{(1)} ( \partial S_0 / \partial y^i,y^i,h_{ij}(t) )$
corresponds to the interaction term of the Hamiltonian, which is
linear in the metric perturbation.

Eq.~(\ref{HJ2}) can be solved by the variable separation method and
the result, which corresponds to the phase of an energy eigenfunction
in the semiclassical approximation, is
\begin{equation}
S_0 (y^i,t) = - E t + 2 \pi \frac{\hbar}{\lambda} n_i y^i
+ \hbar \phi_0
\label{S0},
\end{equation}
where $E$ is an integration constant that corresponds to the energy of
the particle, $\lambda = h (2mE)^{-1/2}$ is the
corresponding de Broglie wavelength, $n^i$ is a unit vector in the
direction of wave propagation and $\phi_0$ is the initial phase at
$t=0$ for any point on the plane defined by $n_i y^i = 0$. The
expression for $S_0 (y^i,t)$ is the same when using either geodesic or
rigid coordinates since the part of the Hamiltonian independent of the
metric perturbation
has the same form in both descriptions. However, one has to take into
account that in general the trajectories of the different elements of
the interferometer, such as mirrors and beam-splitters, will be
different in each coordinate system.

Substituting Eq.~(\ref{S0}) into Eq.~(\ref{HJ3}) we get
\begin{equation}
\frac{d S_1}{d t} =
- H^{(1)} \left( \frac{h}{\lambda} n_i,y^i,h_{ij}(t) \right)
\label{HJ4},
\end{equation}
where $d S_1 / d t = \partial S_1 / \partial t + v^i \partial S_1 /
\partial y^i$ with $v^i = (h / m \lambda) n^i$.
Eq.~(\ref{HJ4}) has the same form for both the description using
geodesic coordinates and that employing rigid coordinates; however,
the explicit expressions for $H^{(1)}$ differ.

The phase difference, up to linear order in the metric perturbation,
between the two ends of the arm of an atomic interferometer of length
$L$ for non-relativistic atoms of energy $E$ propagating along the arm
is $\Delta \phi (t) = \Delta S_0 (t) / \hbar + \Delta S_1 (t) / \hbar$
with
\begin{equation}
\Delta S_0 (t) = S_0 (y^i_{B},t) - S_0 (y^i_{A},t-L/v)
= \frac{h}{\lambda} n_i \left( y^i_{B} (t) - y^i_{A} (t-L/v) \right)
+ E \frac{L}{v}  \label{DS0},
\end{equation}
where $y^i_{A}$ and $y^i_{B}$ are the positions of the two ends of
the arms, which will be different for each coordinate system, and
\begin{equation}
\Delta S_1 (t) = S_1 (y^i_{B},t) - S_1 (y^i_{A},t-L/v)
= - \int\limits^t_{t - L/v} dt' \, H^{(1)} \! \left( \frac{h}{\lambda}
n_i,y^i_{(0)}(t'),h_{ij}(t') \right)
\label{DS1},
\end{equation}
where $y^i_{(0)}(t') = y^i_{A} + (v t') n^i$ and coincides with the
classical trajectory for a particle of mass $m$ and energy $E$ in the
absence of the gravitational wave. Furthermore, we can take the value
for $y^i_{A}$ in either coordinate system because the difference
corresponds to terms that are at least linear in the metric
perturbation and would yield a contribution to Eq.~(\ref{DS1}) of
quadratic or higher order in the metric perturbation.

Eqs.~(\ref{DS0}) and (\ref{DS1}) agree with the general results
obtained in Refs.~\cite{borde01,borde04}.

\subsection{Explicit result using geodesic coordinates}
\label{sec4.2}

If we assume that the arm of the interferometer is perfectly rigid,
\emph{i.e.}, that the physical distance between the two ends remains
constant in time, their positions in rigid coordinates, $x^i_{A}$ and
$x^i_{B}$, are constant\footnote{This is true provided that we take
the origin of coordinates at the center of mass of the interferometer
and it follows a geodesic. In fact, if the gravitational wave
propagates perpendicular to the interferometer, it is sufficient that
its center of mass can move freely along any transverse direction,
\emph{i.e.}, on the plane determined by the two arms.}. Their
positions in geodesic coordinates can be easily obtained, up to linear
order in $h_{ij}$, by inverting Eq.~(\ref{change}):
\begin{eqnarray}
X^i_{A} (t-L/v) &=& x^i_{A} - \frac{1}{2} h^i_j (t - L/v) \, x^j_{A}
\nonumber \\
X^i_{B} (t) &=& x^i_{B} - \frac{1}{2} h^i_j (t) \, x^j_{B}
\label{ends}.
\end{eqnarray}
Thus, in geodesic coordinates, $\Delta S_0 (t)$ is given by
\begin{eqnarray}
\Delta S_0 (t) &=& \frac{h}{\lambda} n_i \left(x^i_{B} -
x^i_{A}\right) - \frac{h}{\lambda} n^i
\left[ \frac{1}{2} h_{ij} (t') y^j_{(0)} (t')\right]_{t' = t - L/v}^{t'=t}
+ E \, \frac{L}{v} \nonumber \\
&=& \frac{h L}{\lambda} - \frac{h}{\lambda} n^i
\left[ \frac{1}{2} h_{ij} (t') y^j_{(0)} (t')\right]_{t' = t - L/v}^{t'=t}
+ E \, \frac{L}{v}
\label{DS0a},
\end{eqnarray}
where $y^j_{(0)} (t) = x^i_{B}$ and $y^j_{(0)} (t - L/v) =
x^i_{A}$, and we took into account that $n_i (x^i_{B} - x^i_{A}) = L$.
Substituting $H^{(1)}$, which is given in this case by the
second term on the right-hand side of Eq.~(\ref{hamiltonian1}), into
Eq.~(\ref{DS1}), one obtains the following result for $\Delta S_1
(t)$:
\begin{equation}
\Delta S_1 (t) = \frac{h v}{2 \lambda} \int \limits ^t _{t - L/v}
dt' h_{ij} (t') n^i n^j
\label{DS1a}.
\end{equation}
Therefore, the total phase difference between the two ends of the
interferometer arm is
\begin{equation}
\Delta \phi (t) = 2 \pi \left( \frac{L}{\lambda} - \frac{n^i}{\lambda}
\left[ \frac{1}{2} h_{ij} (t') y^j_{(0)} (t')\right]_{t' = t - L/v} ^{t' = t}
+ \frac{v}{2 \lambda} \int \limits ^t _{t - L/v} dt' h_{ij} (t') n^i n^j
\right) + \frac{E}{\hbar} \, \frac{L}{v}\label{Dphi1}.
\end{equation}

\subsection{Explicit result using rigid coordinates}
\label{sec4.3}

When rigid coordinates are employed the positions of the arm ends
are fixed at $x_A$ and $x_B$, whence $\Delta S_0 (t)$ is given by
\begin{equation}
\Delta S_0 (t) = \frac{h}{\lambda} n_i
\left(x^i_{B} - x^i_{A}\right) + E \, \frac{L}{v}
= \frac{h L}{\lambda} + E \, \frac{L}{v}
\label{DS0b}.
\end{equation}
Substituting $H^{(1)}$, which corresponds to the second term
on the right-hand side of Eq.~(\ref{hamiltonian2}), into
Eq.~(\ref{DS1}), the result for $\Delta S_1 (t)$ is
\begin{equation}
\Delta S_1 (t) = - \frac{h}{2 \lambda}
\int^t_{t - L/v} dt' \dot{h}_{ij} (t') y^i_{(0)} (t') n^j
= \frac{h v}{2 \lambda} \int^t_{t - L/v} dt' h_{ij} (t') n^i n^j
- \frac{h}{2 \lambda} n^i \left[h_{ij} (t') y^j_{(0)} (t') \right]
^{t'=t} _{t'=t - L/v}
\label{DS1b},
\end{equation}
where we have integrated by parts and taken into account that
$\dot{y}^i_{(0)} (t') = v n^i$. Finally, the total phase difference is
\begin{equation}
\Delta \phi (t) = 2 \pi \left( \frac{L}{\lambda}
- \frac{n^i}{\lambda} \left[ \frac{1}{2} h_{ij} (t') y^j_{(0)} (t')
\right]_{t' = t - L/v} ^{t' = t}
+ \frac{v}{2 \lambda} \int \limits ^t _{t - L/v} dt' h_{ij} (t')
n^i n^j \right) + \frac{E}{\hbar} \, \frac{L}{v}  \label{Dphi2},
\end{equation}
which coincides with Eq.~(\ref{Dphi1}).

The first term on the right-hand side of Eq.~(\ref{Dphi2}), which is
independent of $h_{ij}$ and does not change in time, simply
corresponds to the phase shift in the absence of gravitational waves,
whose effect is contained entirely in the remaining two terms. If we
assume, for simplicity, that the first end of the arm coincides with
the center of mass of the interferometer with rigid arms, and the
interferometer is suspended in such a way that its center of mass can
move freely along the transverse directions (\emph{i.e.}, along any
direction on the plane determined by the two arms), we can take $x_A =
0$. The second term on the right-hand side of Eq.~(\ref{Dphi2}) is
then proportional to $(L / \lambda) n^i n^j h_{ij} (t)$. Finally, if
we consider the particular case of a monochromatic gravitational wave
of angular frequency $\omega$, the last term becomes proportional to
$v / \lambda \omega$ times a dimensionless oscillatory factor,
which is a product of a harmonic function of $\omega t$ and a
harmonic function of $\omega L / v$.

\subsection{Phase shift}
\label{sec4.4}

In the previous subsections we obtained the phase difference between
the two ends of an interferometer arm. Here we explain how to use
those results to compute the phase shift between the two interfering
waves in an interferometer.

First, one divides the trajectory followed by each interfering wave in
stages that describe free propagation between the different elements
of the interferometer: source, beam-splitters, mirrors and
detector. Next, one takes one of the interfering waves and adds the
phase differences for every stage (using Eq.~(\ref{Dphi1})) to obtain
the difference between the phase at the detector location and at a
given observation time $t_\mathrm{f}$, and the phase at the source
location and time $t_\mathrm{f} - \sum_j L_j / v$, where $\sum_j L_j$
is the total length (in the absence of gravitational waves) of the
path followed by that wave. Similarly, one can get the analogous phase
difference for the second interfering wave between the observation
time $t_\mathrm{f}$ and $t_\mathrm{f} - \sum_j L'_j / v$, where
$\sum_j L'_j$ is the total length of the path followed by that wave.
Finally, one needs the phase difference, for a fixed position at the
source location, between times $t_\mathrm{f} - \sum_j L_j / v$ and
$t_\mathrm{f} - \sum_j L'_j / v$. By adding the latter to the phase
difference for the second wave and comparing with the phase difference
for the first wave, one obtains the phase shift between the two
interfering waves.

We now turn to the computation of the phase difference between times
$t_\mathrm{f} - \sum_j L_j / v$ and $t_\mathrm{f} - \sum_j L'_j / v$
at the source location. For simplicity we will consider the situation
in which the source is located at the center of mass of the interferometer
and take into account the remarks made in footnote 2. It is then
natural to assume that the atoms are emitted at constant velocity in
the rigid frame if the size of the source is much smaller than the
arms of the interferometer. This is because the term that accounts for
the interaction with the gravitational wave in rigid coordinates, the
second term on the right-hand side of Eq.~(\ref{hamiltonian2}), is
proportional to the distance from the center of mass and can be
neglected provided that the size of the source is small enough.
Furthermore, under those conditions it is also a good approximation to
assume that atoms emitted at constant velocity in the rigid frame
correspond to atoms emitted at constant total energy $E$ since the
interaction term is negligible. Hence, from Eq.~(\ref{HJ1}) one can
conclude that the phase difference between times $t_\mathrm{f} -
\sum_j L_j$ and $t_\mathrm{f} - \sum_j L'_j / v$ at the source
location is given by
\begin{equation}
\Delta \phi = \frac{E}{\hbar} \left( \sum_j \frac{L'_j}{v}
- \sum_j \frac{L_j}{v} \right)  \label{Dphi3}. 
\end{equation}
When we add this to the phase difference for the second wave and
subtract the result from the phase difference for the first wave to
obtain the phase shift between the two interfering waves, the
contribution from the last term on the right-hand side of
Eqs.~(\ref{Dphi1}) or (\ref{Dphi2}), when added for all the stages of
the path followed by each wave, cancels the right-hand side of
Eq.~(\ref{Dphi3}) exactly. Thus, when computing the phase shift, it is
enough to add the phase differences for each stage keeping only the
first three terms on right-hand side of Eqs.~(\ref{Dphi1}) or
(\ref{Dphi2}).  Note that the whole argument also applies when using
geodesic coordinates, even though the atoms are not emitted at
constant velocity in those coordinates, because the difference between
the Hamiltonians in rigid coordinates and geodesic coordinates is of
the same form as the interaction term in rigid coordinates (see
Eq.~(\ref{change4})).  Therefore, whenever that interaction term can
be neglected for the reasons explained above, one can assume that the
atoms are emitted at constant energy in terms of the geodesic
coordinates as well.

To illustrate how the phase shift is computed with a particular
example, let us consider an interferometer with a Michelson-type
configuration with arms of length $L$ and $L'$, and a negligible
distance from the atom source to the beam-splitter as compared to the
arm lengths. One needs to add the two contributions that correspond to
the two trips in opposite directions along one of the arms using just
the first three terms on the right-hand side of Eqs.~(\ref{Dphi1}) or
(\ref{Dphi2}), proceed analogously for the second arm, and finally
subtract both results. Assuming that the source and beam-splitter are
located at the center of mass and that the interferometer is freely
suspended as specified in footnote 2, the result for the phase shift
is
\begin{eqnarray}
\Delta \phi (t_\mathrm{f}) &=& 2 \pi \left( 2 \frac{L' - L}{\lambda}
- \frac{L'}{\lambda} l^i l^j h_{ij} (t_\mathrm{f} - L'/v)
+ \frac{L}{\lambda} k^i k^j h_{ij} (t_\mathrm{f} - L/v)
+ \frac{v}{2 \lambda} \int \limits ^{t_\mathrm{f}}
_{t_\mathrm{f} - 2L'/v} dt' h_{ij} (t') l^i l^j \right. \nonumber \\
&& \left. - \frac{v}{2 \lambda} \int \limits ^{t_\mathrm{f}}
_{t_\mathrm{f} - 2L/v} dt' h_{ij} (t') k^i k^j \right)
\label{michelson},
\end{eqnarray}
where $k^i$ and $l^i$ are the unit vectors parallel to each arm. For
the particular case in which the two arms are perpendicular and
have equal length and $h_{ij} (t)$ is a harmonic function,
\emph{i.e.}, $h_{ij} (t) = \bar{h}_{ij} \sin (\omega t + \varphi)$,
the previous expression for the phase shift becomes
\begin{eqnarray}
\Delta \phi (t_\mathrm{f})
= 4 \pi k^i k^j \bar{h}_{ij} \left[ \frac{L}{\lambda}
- \frac{v}{\lambda \omega} \sin \left( \frac{\omega L}{v} \right) \right]
\sin \left( \omega t_\mathrm{f} + \varphi  - \frac{\omega L}{v} \right)
\label{michelson2}.
\end{eqnarray}

\subsection{Discrepancy with previous results}
\label{sec4.5}

There is a discrepancy between the result for the phase difference
obtained in Secs.~\ref{sec4.2} and \ref{sec4.3} and that obtained in
Ref.~\cite{chiao04a}, which contains an extra term proportional to
$L^2 \omega / \lambda v$ times a dimensionless oscillatory factor that
dominates in the regime considered there. In this subsection we
explain the relationship between our approach and that of
Refs.~\cite{chiao04a,speliotopoulos04b} for computing the phase
difference between the ends of an interferometer arm. We also point
out a mistake in their analysis which explains the discrepancy with
our result.

The basic quantities in our computation and in that of
Ref.~\cite{chiao04a,speliotopoulos04b} are slightly different. Here we
have considered the solution of the Hamilton-Jacobi equation at fixed
energy to lowest order, whereas in
Ref.~\cite{chiao04a,speliotopoulos04b} the action for a classical
trajectory that goes from $x_A$ to $x_B$ in a given time was employed.
Although these two objects share some common properties (they both
satisfy the Hamilton-Jacobi equation~(\ref{HJ1})) they have different
meanings. This point can be easily illustrated by considering the
example of a one-dimensional time-independent Hamiltonian. The analog
of the object that we considered is a solution of the Hamilton-Jacobi
equation of fixed energy $E$, \emph{i.e.}, a function $S_E(x,t)$ which
satisfies the Hamilton-Jacobi equation and such that $\partial
S_E(x,t) / \partial t = - E$. On the other hand, the analog of the
object considered in Ref.~\cite{chiao04a,speliotopoulos04b} is the
function $S(x,x',t)$ which results from evaluating the classical
action for a classical trajectory that goes from $x'$ to $x$ in a
period of time $t$. It can also be obtained from $S_E(x,t)$ by
choosing the energy of the classical trajectory characterized by $x'$,
$x$ and $t$: $S(x,x',t) = S_{E(x,x',t)}(x,t)$. Although one can show
that $S(x,x',t)$ also satisfies the Hamilton-Jacobi equation,
$S_E(x,t)$ and $S(x,x',t)$ are not equivalent as functions of $x$ and
$t$. Furthermore, whereas $S_E(x,t)$ can be identified with the
exponent of the lowest order JWKB approximation for an
\emph{eigenfunction} of energy $E$:
\begin{equation}
\psi_E (x,t) \sim e^{iS_E(x,t) / \hbar}
\label{eigenfunc},
\end{equation}
$S(x,x',t)$ corresponds to the exponent of the lowest order
semiclassical approximation of the \emph{propagator}:
\begin{equation}
K(x,x',t) = \sum_E \psi_E (x,t) \psi^*_E (x',0) \sim e^{iS(x,x',t)/\hbar}
\label{propag},
\end{equation}
where the sum is over all the energy eigenvalues, including possible
degeneracies, associated with the Hamiltonian that one is considering.
Both $\psi_E (x,t)$ and $K(x,x',t)$ satisfy the same time dependent
Schr\"odinger equation, but they are again different functions of
$x$ and $t$.  Since the physical situation that we are interested in
is the interference of atoms emitted with a fixed energy (up to a good
approximation) propagating along the arms of the interferometer,
$\psi_E (x,t)$ (and $S_E(x,t)$ in the semiclassical approximation) is
the relevant quantity that should be employed to obtain the phase
difference directly.

Nevertheless, one can still make use of $S(x,x',t)$ to compute phase
differences for an interferometer. This follows from the observation
that by evaluating the action for a classical trajectory of energy $E$
that goes from $x$ to $x'$ one gets $S(x,x',T) = S_E(x',t) - S_E(x,t -
T)$, where $T$ is the time needed to go from $x$ to $x'$. The phase
$S_E(x_\mathrm{f},t)$ at the detector location for one of the
interfering waves in an interferometer computed in this way is given
by the action for the whole trajectory that corresponds to the path
followed by that wave in the interferometer plus $S_E(x_\mathrm{i},t -
T)$, where $x_\mathrm{i}$ is the position of the atom source and $T$
is the time the whole trajectory spends in the interferometer. The
phase shift between the two interfering waves is then given by the
difference of the actions for the classical trajectories that
correspond to the two paths followed by the two interfering waves plus
$S_E(x_\mathrm{i},t - T_2) - S_E(x_\mathrm{i},t - T_1)$, where $T_1$
and $T_2$ are the times spent by the classical trajectories in the
interferometer. For a time-independent Hamiltonian the last
contribution simply corresponds to $E (T_2 - T_1)$.

A similar procedure can be employed when a time-dependent perturbation
is included in the Hamiltonian. In this case one needs to evaluate
perturbatively the action for the classical trajectory associated with
each interfering wave, which is also determined perturbatively. When
computing the phase difference between the two ends of an
interferometer arm, one should consider solutions of the perturbed
equations of motion that go from one end of the arm to the
other. Furthermore, it should also be taken into account that the
kinematic momentum is reflected on the mirrors and, when using rigid
coordinates, the atoms are emitted with constant velocity, as
explained in Sec.~\ref{sec4.4}. For a given observation time at the
detector, the previous conditions completely determine the
trajectory. That means that the initial time at which that trajectory
leaves the detector will in general be different from the initial time
for the unperturbed trajectory.

In Appendix~\ref{appA} we derive the following useful formula for the
perturbed action associated with a perturbed classical solution that
goes from one end of an interferometer arm to the other:
\begin{equation}
S_0 [x_0(t') + x_1(t')] + S_1 [x_0(t')]
= \int^{t_B}_{t_A} dt' L^{(0)}(x_0(t'))
- E (\Delta t_B - \Delta t_A)
-  \int^{t_B}_{t_A} dt' H^{(1)} (x_0(t'),p_0(t'))
\label{int4b},
\end{equation}
with $t_B = t$ and $t_A = t - L / v$, where $L$ is
the length of the arm. The trajectory $x_0(t')$ is the solution of the
classical equation of motion associated with the unperturbed
Lagrangian $L^{(0)}$ that goes from one end at
time $t_A$ to the other end at time $t_B$, and
$p_0(t')$ is the unperturbed canonical momentum associated with the
trajectory $x_0(t')$ at the time $t'$. The perturbation $x_1(t')$
vanishes at the perturbed initial and final times. 
The contribution from the lowest order term,
the first term on the right-hand side of Eq.~(\ref{int4b}), is
identical to the time independent case and coincides with the result
in Eq.~(\ref{DS0b}). The last term on the right-hand side of
Eq.~(\ref{int4b}) in turn coincides with the result in Eq.~(\ref{DS1b}).
Thus, when we add the phase differences for every stage obtained
with Eq.~(\ref{int4b}), we recover the same result as in Sec.~\ref{sec4.4}
except for the contributions that come from the second term on the
right-hand side of Eq.~(\ref{int4b}). Let us look at those more
carefully. We are considering a given observation time at the detector
location, hence $\Delta t_B$ vanishes for the last
stage. Moreover, the initial time perturbation $\Delta t_A$
for any stage is always canceled by the final time perturbation
$\Delta t_B$ for the previous stage. Altogether we are left
only with the term that corresponds to the initial time perturbation
at the source location, $\Delta t_\mathrm{i}$, for the first
stage. This gives rise to a term $(E / \hbar) (\Delta t_\mathrm{i}
- \Delta t'_\mathrm{i})$ when subtracting the total phase
difference for the two interfering waves. On the other hand, one also
needs to include the phase difference at the source location between
the initial times $t_\mathrm{i} + \Delta t_\mathrm{i}$ and
$t'_\mathrm{i} + \Delta t'_\mathrm{i}$, which corresponds to the
trajectories associated with the first and second interfering wave
respectively. From the explanation provided in Sec.~\ref{sec4.4} it
follows that such a phase difference is given in the present case by
\begin{equation}
\Delta \phi = \frac{E}{\hbar} \left( t_\mathrm{i}
+ \Delta t_\mathrm{i} - t'_\mathrm{i} - \Delta t'_\mathrm{i} \right)
\label{source} ,
\end{equation}
and the terms involving $\Delta t_\mathrm{i}$ and $\Delta
t'_\mathrm{i}$ cancel exactly those coming from the phase difference
for the two interfering waves when computing the phase shift between
them. Therefore, the result for the phase
shift is entirely equivalent to that obtained by following the
procedure based on the Hamilton-Jacobi equation employed in
Secs.~\ref{sec4.1}-\ref{sec4.4}.

Chiao and Speliotopoulos obtained a different result in
Refs.~\cite{chiao04a,speliotopoulos04b} because they fixed the initial
and final times for each stage to the same value as in the unperturbed
case. Since they also considered a fixed initial velocity and
position, the perturbed trajectory no longer coincide with the
detector location at final time, but the authors did not realize that
they were implicitly using a perturbed trajectory with the wrong
boundary condition (additional details on this point can be found in
Appendix~\ref{appA}). As explained in the previous paragraph, when
the appropriate boundary conditions are employed, including perturbed
initial and final times for each stage, the result obtained in
Sec.~\ref{sec4.4} is recovered.

\section{Discussion}
\label{sec5}

We divide our discussion into two parts. First, we will examine the
physical implications of the result found in the previous section.
Next, we will discuss some fundamental aspects about the use of atom
interferometers for the detection of gravitational waves. We will not
assume any specific scheme for the interferometer (\emph{e.g.} its
geometry), but keep the discussion general by addressing the main
qualitative features independent of the particular configuration.

The phase shift between the two interfering waves in an interferometer
is obtained by adding the phase difference (obtained in
Sec.~\ref{sec4}) between the ends of every arm and comparing the
result for the two trajectories that correspond to the two interfering
waves.  Since the phase difference is given by an expression of the
same form for all arms, the phase shift between the two waves will
also have three contributions qualitatively similar to the first three
terms in Eq.~(\ref{Dphi1}). The first contribution is proportional to
the difference between the total length of the paths for the two
interfering waves and is the only term in the absence of gravitational
waves. The third contribution encodes the entire effect of the
gravitational wave on the propagation of the atoms if the mirrors were
freely suspended. Finally, the second contribution accounts for the
fact that we are considering an interferometer with rigid arms.

One can consider two regimes depending on the ratio of the transit
time of the atoms to the period of the gravitational wave. We will
refer to the case in which the period of the gravitational wave is
smaller than the transit time of the atoms along the arms of the
interferometer as the \emph{high frequency regime}. In that regime the
second contribution mentioned above dominates. Correspondingly, the
\emph{low frequency regime} refers to the case in which the period of
the gravitational wave is greater than the transit time of the atoms.
In that case the second and third contributions become comparable, but
since they carry opposite signs their effects cancel out.

Despite the obvious differences between atom and light
interferometers, \emph{e.g.}, the atoms in MIGO are non-relativistic,
whereas photons in LIGO are highly relativistic, the expressions for
the phase shift in the two cases are formally equivalent, except for
the second contribution, provided that one 1) replaces the de Broglie
wavelength $\lambda$ with the laser wavelength $\lambda_\mathrm{ph}$
and $v$ with $c$ (the speed of light), 2) uses the effective length
for the arms of the laser interferometer obtained by multiplying the
arm length by the number of times photons travel back and forth inside
the Fabry-Perot cavities, $b$. The second contribution mentioned above
is absent for LIGO because its mirrors are freely suspended. If its
arms were rigid, the whole expression would be equivalent for atom and
laser interferometers.

Laser interferometers with freely suspended mirrors such as LIGO are
mostly sensitive in the low frequency regime, where the expression for
the phase shift becomes $(b \, L / \lambda_\mathrm{ph}) h_{ij} n^i
n^j$ times an oscillatory factor of order one which is a harmonic
function of $\omega t$, where $t$ is the observation time and
$\omega$ is the angular frequency of the gravitational wave. On the
other hand, for high frequency gravitational waves the phase shift
becomes proportional to $(c / \omega \lambda_\mathrm{ph}) h_{ij} n^i
n^j$ times $\sin (\omega b L/c)$ and a harmonic function of
$\omega t$. Therefore, the phase shift in the high frequency
regime is reduced by an amount which corresponds to replacing the
effective length of the arm with the distance traveled by the photons
during one gravitational wave period, which is smaller, in that
regime, than the effective length of the arms. This can be
qualitatively understood as follows: when several oscillations of the
gravitational wave take place while each photon travels along the
interferometer arms, the effect is averaged out and only the last
cycle of the gravitational wave counts.

In contrast, for most conceivable situations, MIGO would typically be
implemented in the high frequency regime, since $v \ll c$ for
non-relativistic atoms and the transit time becomes much larger than
in laser interferometers. In that regime, the second contribution
dominates and the phase shift becomes $L / \lambda$ times the
gravitational wave amplitude and a harmonic function of $\omega t$.
MIGO is not affected by that averaging-out effect for high frequencies
as laser interferometers because its arms are rigid. Thus, it is
better to consider rigid arms for MIGO rather than freely suspended
mirrors because it works in the high frequency regime\footnote{In a
spacetime corresponding to a harmonic linearized gravitational wave
with wavelength $\lambda_\mathrm{GW}$ it is only possible to have
rigid objects with a size along the transverse direction of the order
of $\lambda_\mathrm{GW}$ at most. For an atom interferometer it is
possible to fulfill that condition and still be in the high frequency
regime because $v \ll c$. On the other hand, for a laser
interferometer that condition can only be satisfied while being in the
high frequency regime provided that the photons bounce back and forth
many times inside the Fabry-Perot cavities.}. On the other hand, for
low frequencies the second and third terms become comparable
but with opposite sign and cancel out. This is the same reason why
LIGO would not work if its mirrors were not freely suspended.

However, even though an atom interferometer with rigid arms would not
exhibit the averaging-out effect mentioned above, the oscillatory
factor limits the amount of time during which the measurement can be
performed. For a fixed flux of atoms, that means that, as the
gravitational wave frequency increases, the maximum value for the
total number of atoms involved in the measurement decreases and shot
noise becomes important, eventually limiting the sensitivity of the
interferometer.

Once the correct result for the phase shift, obtained by applying
Eq.~(\ref{Dphi1}) to the particular configuration of interest, is
employed, the major claims by Chiao and Speliotopoulos about the
capabilities of MIGO no longer hold. First, one can immediately see
from Eq.~(39) that it is better to use fast atoms than slow ones.
Second, although the averaging-out effect is not so drastic as for
laser interferometers, the phase shift does not increase with the
gravitational wave frequency. In fact, if one considers gravitational
waves which correspond to the low frequency regime for LIGO, the
expression for the phase shift is given by $b L_\mathrm{LIGO} /
\lambda_\mathrm{ph}$ times the amplitude of the gravitational wave and
a harmonic function of $\omega t$. On the other hand, that would
correspond to the high frequency regime for MIGO in most conceivable
situations, and the phase shift would be $L_\mathrm{MIGO} / \lambda$
times the amplitude of the gravitational wave and a similar harmonic
function of $\omega t$. Thus, under equal conditions (\emph{i.e.}, if
one had $\lambda = \lambda_\mathrm{ph}$, $L_\mathrm{MIGO} =
L_\mathrm{LIGO}$ and assumed in addition that the atoms could bounce
back and forth the same number of times) the sensitivities of MIGO and
LIGO would be comparable, so that the great enhancement purported in
Ref.~\cite{chiao04a} does not really exist.

The assumptions that $L_\mathrm{MIGO} = L_\mathrm{LIGO}$ and the atoms
can bounce a similar number of times as the photons in the Fabry-Perot
cavities are of course unrealistic. Nevertheless, one might still
wonder whether a feasible range of parameters exists for which the
sensitivity of MIGO could be comparable to that of laser
interferometers. The main advantage of atom interferometers is that
much shorter wavelengths can be used, whereas photons with much
shorter wavelength would be difficult to work with. In theory, this
could make it possible to decrease the size of the interferometer to
more realistic values. In practice, however, there are a number of
noise sources involving the position of the mirrors which cannot be
overcome by increasing the idealized sensitivity of the
interferometer, so that the displacement induced by the gravitational
wave needs to be above a certain minimum value, which implies a
minimum size for the interferometer. In fact, that is the reason why
the size of LIGO could not be reduced simply by using shorter
wavelength photons, since its sensitivity at low and intermediate
frequencies is limited by seismic and suspension thermal noise. The
fact that MIGO has rigid arms rather than freely suspended mirrors
does not seem to improve the situation drastically. Finally, as should
be quite obvious from the beginning, the major limitation for an atom
interferometer is the maximum flux of atoms that can be produced as
compared to the huge flux of photons available in a laser
interferometer. Therefore, for realistic situations MIGO would likely
be limited by shot noise.

We close this section by pointing out that our conclusions are
essentially the same both for interferometers which have a spatial
projection of their arms with non-vanishing area and for
interferometers with a Michelson-type configuration such as
LIGO. According to Ref.~\cite{chiao03}, this distinction is crucial
for static spacetimes with nontrivial curvature, because only the
former type can be employed to measure the spacetime
curvature. Nevertheless, it is not so important for linearized
gravitational waves propagating on a Minkowski background. This point
bears strongly on the claims made in Ref.~\cite{speliotopoulos04b}
which depend on selecting the right expression for the phase shift
among the different results obtained there. The authors chose the
expression that coincides with the result for the static case when the
static limit of the gravitational wave is considered. However, the
static limit of linearized gravitational waves (propagating on a
Minkowski background) is flat spacetime itself and, hence, its
curvature vanishes. The static limit is trivial and cannot be used for
that purpose, whereas for non-vanishing gravitational wave
frequencies, even if they are small, the result of Ref.~\cite{chiao03}
for the static case is no longer valid. Anyway, when the correct
procedure is employed to compute the phase shift, the result is the
same for different coordinate systems, as one could anticipate from
the principle of general covariance. To begin with, there is no need
to find a criterion for selecting the right result.


\begin{acknowledgments}
We thank Ho Jung Paik for interesting comments.  We also acknowledge a
number of discussions with Raymond Chiao and Achilles Speliotopoulos,
and thank them for their interest in our approach.  B.~L.~H.\ and
A.~R.\ are supported by NSF under Grant PHY03-00710, and W.~D.~P.\ is
supported by the ONR and by NASA.
\end{acknowledgments}

\appendix

\section{Action for perturbed trajectories around a classical solution}
\label{appA}

In Sec.~\ref{sec4.5} we explained that one can obtain the phase
difference between the two ends of an interferometer arm at times
$t_A + \Delta t_A$ and $t_B + \Delta
t_B$ by evaluating the action for the perturbed trajectory
that satisfies the perturbed equation of motion and goes from one end
at time $t_A + \Delta t_A$ to the other end at time
$t_B + \Delta t_B$. Up to first order in the
perturbation, the action is given by
\begin{equation}
S_0 [x_0(t') + x_1(t')] + S_1 [x_0(t')]
= \int^{t_B + \Delta t_B}_{t_A + \Delta
t_A} dt' L^{(0)}(x_0(t') + x_1(t'))
+ \int^{t_B}_{t_A} dt' L^{(1)}(x_0(t'))
\label{int4c},
\end{equation}
where $x_0(t')$ is the solution of the unperturbed equation of motion
that goes from one end at time $t_A$ to the other end at time
$t_B$, and $x_1(t')$ is the perturbation of the trajectory. We
have assumed that the perturbations of the initial and final times are
of the same order as $x_1(t')$ and the Lagrangian perturbation
$L^{(1)}$.  In this Appendix we will derive a useful expression for
that action. In doing so, we will also point out the mistake made in
Refs.~\cite{chiao04a,speliotopoulos04b}.

Let us start with the action for a general Lagrangian $L$
\begin{equation}
S[x(t)] = \int ^{t_B} _{t_A} L (x(t)) dt , 
\end{equation}
where $x(t)$ is a solution of the classical equation of motion
associated with the Lagrangian $L$, and add a small perturbation
$\delta x (t)$ to $x(t)$. The usual variational treatment yields the
following result up to linear order in $\delta x (t)$:
\begin{equation}
\delta S = p_B \delta x_B - p_A \delta x_A ,
\end{equation}
where the subindices $A$ and $B$ denote quantities
evaluated at the initial and final time respectively. On the other
hand, if we perturb the initial and final times, the perturbed action
\begin{equation}
\int ^{t_B + \Delta t_B} _{t_A + \Delta t_A} L (x(t)) dt
\end{equation}
minus the original action, up to linear order in $\Delta t_A$
and $\Delta t_B$, becomes $L (x(t_B)) \Delta
t_B - L (x(t_A)) \Delta t_A$. When both
kinds of perturbations are considered simultaneously, the final
position changes, up to linear order, by $\Delta x_B = \delta
x_B + \dot{x}_B \Delta t_B$ (considering that
$\delta x_B$ and $\dot{x}_B \Delta t_B$ are
of the same order). A similar result also holds for the initial time.
Altogether the change in the action is given by:
\begin{equation}
\delta S = \left( p_B \Delta x_B - p_B \dot{x}_B \Delta t_B + L (x(t_B))
\Delta t_B \right)
 - \left( p_A \Delta x_A - p_A \dot{x}_A \Delta t_A + L (x(t_A))
\Delta t_A \right) ,
\end{equation}
which can be equivalently rewritten as
\begin{equation}
\delta S = \left( p_B \Delta x_B - H(x_B,p_B) \Delta t_B \right)
 - \left( p_A \Delta x_A - H(x_A,p_A) \Delta t_A \right)
\label{deltaS}.
\end{equation}

Applying Eq.~(\ref{deltaS}) to the case in which the Lagrangian is
$L^{(0)}$ and the perturbation is $\delta x (t) = x_1 (t)$, we get
\begin{equation}
\delta S_0 = \left( p_B \Delta x_B - p_A \Delta x_A \right)
 - \left( H^{(0)}(x_B,p_B) \Delta t_B - H^{(0)}(x_A,p_A) \Delta t_A \right)
\label{deltaS2},
\end{equation}
where $p_A$ and $p_B$ are the momenta at the initial
and final time for the unperturbed trajectory. Since the emission and
observation points are fixed (when using rigid coordinates), we should
take $\Delta x_A = \Delta x_B = 0$. Hence, the first
term on the right-hand side of Eq.~(\ref{int4c}) can be rewritten as
\begin{equation}
 \int^{t_B}_{t_A} dt' L^{(0)} (x_0(t'))  - \left( H^{(0)} (x_B,p_B)
\Delta t_B - H^{(0)}(x_A,p_A) \Delta t_A \right)
=  \int^{t_B}_{t_A} dt' L^{(0)}(x_0(t')) - E (\Delta t_B - \Delta t_A) 
\label{int3},
\end{equation}
where we have taken into account that $x_0(t)$ is a constant-energy
solution of the unperturbed equation of motion with energy $E$.

In addition, one can show that $L^{(1)} (x_0(x,p),\dot{x}_0(x,p)) =
- H^{(1)} (x,p)$.  This can be seen as follows. Consider the relation
between momentum and velocity for the perturbed
Lagrangian\footnote{From now on in this Appendix, all the equalities should
be regarded as valid only up to linear order in the perturbation.}: $\dot{x}
(x,p) = \dot{x}_0 (x,p) + \dot{x}_1 (x,p)$, where $\dot{x}_0 (x,p)$
gives the velocity in terms of the momentum when only the unperturbed
Lagrangian is considered. Taking this into account, we have
\begin{equation}
L^{(0)}(x,\dot{x}(x,p)) = L^{(0)}(x,\dot{x}_0(x,p)) + \left( \frac{\partial
L^{(0)}(x,\dot{x}_0)}{\partial \dot{x}_0} \right) \! (x,p)\, \dot{x}_1 (x,p)
= L^{(0)}(x,\dot{x}_0(x,p)) + p \, \dot{x}_1 (x,p) .
\end{equation}
We can then obtain the Hamiltonian:
\begin{eqnarray}
H(x,p) &=& p (\dot{x}_0(x,p) + \dot{x}_1(x,p)) - L^{(0)}(x,\dot{x}(x,p))
- L^{(1)}(x,\dot{x}(x,p)) \nonumber \\
&=& p (\dot{x}_0 + \dot{x}_1) - L^{(0)}(x,\dot{x}_0(x,p)) - p \dot{x}_1
- L^{(1)} = H^{(0)} (x,p) - L^{(1)}(x,p) ,
\end{eqnarray}
and verify that $H^{(1)} (x,p) = - L^{(1)} (x_0(x,p),\dot{x}_0(x,p))$.
Together with Eq.~(\ref{int3}) this implies that Eq.~(\ref{int4c}) can
be re-expressed as
\begin{equation}
S_0 [x_0(t') + x_1(t')] + S_1 [x_0(t')]
=  \int^{t_B}_{t_A} dt' L^{(0)}(x_0(t'))
- E (\Delta t_B - \Delta t_A)
-  \int^{t_B}_{t_A} dt' H^{(1)} (x_0(t'),p_0(t'))
\label{int4d}.
\end{equation}

In Refs.~\cite{chiao04a,speliotopoulos04b}, the authors always
considered $\Delta t_A = \Delta t_B = 0$. However, since the atoms are
emitted at constant velocity (in rigid coordinates), in general one
will have $\Delta x_\mathrm{B} \neq 0$ for the first stage, and
similarly for the remaining stages. This violates the boundary
conditions that should be satisfied when computing the phase
difference between the ends of an interferometer arm: $\Delta
x_\mathrm{A} = \Delta x_\mathrm{B} = 0$. Furthermore, when evaluating
the action for the perturbed trajectory under this conditions, the
term $- E (\Delta t_B - \Delta t_A)$ in Eq.~(\ref{int4d}) is actually
replaced by $p_B \Delta x_B - p_A \Delta x_A$. One can explicitly
check that this indeed corresponds to the extra term obtained in
Refs.~\cite{chiao04a,speliotopoulos04b} as compared to our result.

\section{Possible effect due to the reflection off the mirrors}
\label{appB}

In this Appendix we explain why, contrary to the claim in
Ref.~\cite{foffa04b}, the possible effects due to the reflection off
the mirrors cancel out in the final result for the phase shift.

It should be stressed that, when computing the phase difference for
each arm, any effects due to the change of the classical trajectories
as a result of the reflection off the mirror are already included in
the treatment provided in Appendix~\ref{appA}. Indeed, the results
there are valid for an arbitrary small perturbation of the classical
trajectory employed to compute the phase difference. Therefore, it is
clear from Eq.~(\ref{int4d}) that the only contribution that follows
from integrating the unperturbed Lagrangian along the perturbed
trajectory is $- E (\Delta t_B - \Delta t_A)$. Note that we can take
$\Delta x_\mathrm{A} = \Delta x_\mathrm{B} = 0$ not only for rigid
coordinates but also for geodesic coordinates. The position of the
mirror in geodesic coordinates for a rigid arm is given by
Eq.~(\ref{ends}) and, although the time shifts $\Delta t_A$ and
$\Delta t_B$ can give rise to non-vanishing $\Delta x_\mathrm{A}$ and
$\Delta x_\mathrm{B}$, those contributions are of higher order in
$h_{ij}$, because $\Delta t_A$ and $\Delta t_B$ are already of order
$h_{ij}$, and can be neglected.

\begin{figure}
\includegraphics[height=8cm]{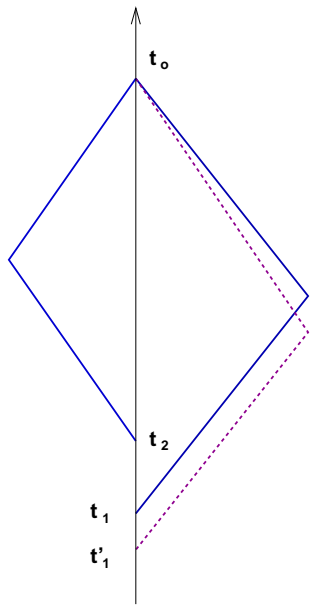}
\caption[fig1]{Schematic spacetime diagram representing the two
trajectories employed to compute the phase shift in an atomic
interferometer with a Michelson-type configuration. The vertical axis
corresponds to time and the horizontal axis represents the positions
along the two arms. Each side of the time axis represents a different
direction in space which corresponds to each one of the arms. The
continuous lines correspond to the unperturbed trajectories, and the
dashed line to a perturbed one. In this particular example the
perturbed trajectory is identical to the unperturbed one before
bouncing off the mirror, but its velocity is higher after the reflection.}
\label{fig1}
\end{figure}

In fact, one can easily check that Eq.~(\ref{int4d}) is equivalent to
Eqs.~(4) or (5) in Ref.~\cite{foffa04b}. However, the authors of
Ref.~\cite{foffa04b} did not realize that the contributions $- E
(\Delta t_B - \Delta t_A)$ for each piece of the trajectory cancel out
in the final result for the phase shift. This follows from the
discussion in the next to last paragraph of Sec.~\ref{sec4.5}. For a
fixed observation time, adding those contributions for the several
pieces of the trajectory associated with one of the interfering waves
gives $E \Delta t_\mathrm{i}$, where $\Delta t_\mathrm{i}$ is the
shift of the time at which the perturbed classical trajectory leaves
the source as compared to the unperturbed one. An analogous result, $E
\Delta t'_\mathrm{i}$, is obtained for the other interfering wave.
These two contributions are, nevertheless, canceled by the
contribution from the phase difference at the source location, which
is given by Eq.~(\ref{source}). This is schematically represented in
Fig.~\ref{fig1}, where only one of the perturbed trajectories is
shown. In order to compute the phase shift at the observation time
$t_\mathrm{o}$, one needs to evaluate the action for the classical
trajectory that leaves the source at time $t_1$, is reflected off the
mirror and reaches the detector at $t_\mathrm{o}$, and then compare it
with the result of adding the action for the trajectory that leaves at
$t_2$ to the phase difference between $t_1$ and $t_2$ at the source
location. When a perturbed trajectory is considered (dashed line), the
change in the action for that trajectory is exactly canceled by the
phase difference between $t'_1$ and $t_1$ at the source location.
Thus, it is actually irrelevant whether the perturbed trajectory or
the unperturbed one is used when computing the phase shift.

We conclude this Appendix with several remarks. First, our previous
discussion only makes sense if the difference between the initial
times for the two trajectories is smaller than the coherence time of
the wave-packets emitted by the source, but this is a necessary
condition for the interferometer to work. Second, if the contribution
due to the phase difference at the source location is not taken into
account, one does not get the right result for the phase shift in an
interferometer with different arm lengths even in the absence of
gravitational waves (the contributions from the last term in
Eqs.~(\ref{Dphi1}) or (\ref{Dphi2}) would not be canceled by the
source contribution given by Eq.~(\ref{Dphi3})). In fact, for a laser
interferometer the phase shift obtained by considering light rays in
the geometrical optics approximation would be entirely given by the
difference of times at which the light rays leave the source. Finally,
instead of considering the phase difference between different times at
the same location for the source, one can alternatively consider
classical trajectories with the same initial time but slightly
different initial positions and include the phase difference between
those two positions for the initial state (the wavefunction at that
initial time). This approach was employed in Ref.~\cite{storey94} and
gives an equivalent result\footnote{We thank Raymond Chiao and
Achilles Speliotopoulos for pointing out Ref.~\cite{storey94} to us.}.


\begin{thebibliography}{15}
\expandafter\ifx\csname natexlab\endcsname\relax\def\natexlab#1{#1}\fi
\expandafter\ifx\csname bibnamefont\endcsname\relax
  \def\bibnamefont#1{#1}\fi
\expandafter\ifx\csname bibfnamefont\endcsname\relax
  \def\bibfnamefont#1{#1}\fi
\expandafter\ifx\csname citenamefont\endcsname\relax
  \def\citenamefont#1{#1}\fi
\expandafter\ifx\csname url\endcsname\relax
  \def\url#1{\texttt{#1}}\fi
\expandafter\ifx\csname urlprefix\endcsname\relax\def\urlprefix{URL }\fi
\providecommand{\bibinfo}[2]{#2}
\providecommand{\eprint}[2][]{\url{#2}}

\bibitem[{\citenamefont{Chiao and Speliotopoulos}(2004)}]{chiao04a}
\bibinfo{author}{\bibfnamefont{R.~Y.} \bibnamefont{Chiao}} \bibnamefont{and}
  \bibinfo{author}{\bibfnamefont{A.~D.} \bibnamefont{Speliotopoulos}},
  \bibinfo{journal}{J. Mod. Opt.} \textbf{\bibinfo{volume}{51}},
  \bibinfo{pages}{861} (\bibinfo{year}{2004}).

\bibitem[{\citenamefont{Foffa et~al.}(2004{\natexlab{a}})\citenamefont{Foffa,
  Gasparini, Papucci, and Sturani}}]{foffa04a}
\bibinfo{author}{\bibfnamefont{S.}~\bibnamefont{Foffa}},
  \bibinfo{author}{\bibfnamefont{A.}~\bibnamefont{Gasparini}},
  \bibinfo{author}{\bibfnamefont{M.}~\bibnamefont{Papucci}}, \bibnamefont{and}
  \bibinfo{author}{\bibfnamefont{R.}~\bibnamefont{Sturani}}
  (\bibinfo{year}{2004}{\natexlab{a}}), \eprint{gr-qc/0407039}.

\bibitem[{\citenamefont{Speliotopoulos and
  Chiao}(2004{\natexlab{a}})}]{speliotopoulos04a}
\bibinfo{author}{\bibfnamefont{A.~D.} \bibnamefont{Speliotopoulos}}
  \bibnamefont{and} \bibinfo{author}{\bibfnamefont{R.~Y.} \bibnamefont{Chiao}},
  \bibinfo{journal}{Phys. Rev. D} \textbf{\bibinfo{volume}{69}},
  \bibinfo{pages}{084013} (\bibinfo{year}{2004}{\natexlab{a}}).

\bibitem[{\citenamefont{Linet and Tourrenc}(1976)}]{linet76}
\bibinfo{author}{\bibfnamefont{B.}~\bibnamefont{Linet}} \bibnamefont{and}
  \bibinfo{author}{\bibfnamefont{P.}~\bibnamefont{Tourrenc}},
  \bibinfo{journal}{Can. J. Phys.} \textbf{\bibinfo{volume}{54}},
  \bibinfo{pages}{1129} (\bibinfo{year}{1976}).

\bibitem[{\citenamefont{Stodolsky}(1979)}]{stodolsky79}
\bibinfo{author}{\bibfnamefont{L.}~\bibnamefont{Stodolsky}},
  \bibinfo{journal}{Gen. Relativ. Gravit.} \textbf{\bibinfo{volume}{11}},
  \bibinfo{pages}{391} (\bibinfo{year}{1979}).

\bibitem[{\citenamefont{Cai and Papini}(1989)}]{cai89}
\bibinfo{author}{\bibfnamefont{Y.~Q.} \bibnamefont{Cai}} \bibnamefont{and}
  \bibinfo{author}{\bibfnamefont{G.}~\bibnamefont{Papini}},
  \bibinfo{journal}{Class. Quant. Grav.} \textbf{\bibinfo{volume}{6}},
  \bibinfo{pages}{407} (\bibinfo{year}{1989}).

\bibitem[{\citenamefont{Bord\'e}(2001)}]{borde01}
\bibinfo{author}{\bibfnamefont{C.~J.} \bibnamefont{Bord\'e}},
  \bibinfo{journal}{C. R. Acad. Sci. Paris (S\'erie IV)}
  \textbf{\bibinfo{volume}{2}}, \bibinfo{pages}{509} (\bibinfo{year}{2001}).

\bibitem[{\citenamefont{Bord\'e}(2004)}]{borde04}
\bibinfo{author}{\bibfnamefont{C.~J.} \bibnamefont{Bord\'e}},
  \bibinfo{journal}{Gen. Relativ. Gravit.} \textbf{\bibinfo{volume}{36}},
  \bibinfo{pages}{475} (\bibinfo{year}{2004}).

\bibitem[{\citenamefont{Speliotopoulos and
  Chiao}(2004{\natexlab{b}})}]{speliotopoulos04b}
\bibinfo{author}{\bibfnamefont{A.~D.} \bibnamefont{Speliotopoulos}}
  \bibnamefont{and} \bibinfo{author}{\bibfnamefont{R.~Y.} \bibnamefont{Chiao}}
  (\bibinfo{year}{2004}{\natexlab{b}}), \eprint{gr-qc/0406096}.

\bibitem[{\citenamefont{Foffa et~al.}(2004{\natexlab{b}})\citenamefont{Foffa,
  Papucci, and Sturani}}]{foffa04b}
\bibinfo{author}{\bibfnamefont{S.}~\bibnamefont{Foffa}},
  \bibinfo{author}{\bibfnamefont{M.}~\bibnamefont{Papucci}}, \bibnamefont{and}
  \bibinfo{author}{\bibfnamefont{R.}~\bibnamefont{Sturani}}
  (\bibinfo{year}{2004}{\natexlab{b}}), \eprint{gr-qc/0409099}.

\bibitem[{\citenamefont{Misner et~al.}(1973)\citenamefont{Misner, Thorne, and
  Wheeler}}]{misner73}
\bibinfo{author}{\bibfnamefont{C.~W.} \bibnamefont{Misner}},
  \bibinfo{author}{\bibfnamefont{K.~S.} \bibnamefont{Thorne}},
  \bibnamefont{and} \bibinfo{author}{\bibfnamefont{J.~A.}
  \bibnamefont{Wheeler}}, \emph{\bibinfo{title}{Gravitation}}
  (\bibinfo{publisher}{Freeman}, \bibinfo{address}{San Francisco},
  \bibinfo{year}{1973}).

\bibitem[{\citenamefont{Wald}(1994)}]{wald94}
\bibinfo{author}{\bibfnamefont{R.~M.} \bibnamefont{Wald}},
  \emph{\bibinfo{title}{Quantum field theory in curved spacetime and black hole
  thermodynamics}} (\bibinfo{publisher}{The University of Chicago Press},
  \bibinfo{address}{Chicago}, \bibinfo{year}{1994}).

\bibitem[{\citenamefont{Speliotopoulos}(1995)}]{speliotopoulos95}
\bibinfo{author}{\bibfnamefont{A.~D.} \bibnamefont{Speliotopoulos}},
  \bibinfo{journal}{Phys. Rev. D} \textbf{\bibinfo{volume}{51}},
  \bibinfo{pages}{1701} (\bibinfo{year}{1995}).

\bibitem[{\citenamefont{Chiao and Speliotopoulos}(2003)}]{chiao03}
\bibinfo{author}{\bibfnamefont{R.~Y.} \bibnamefont{Chiao}} \bibnamefont{and}
  \bibinfo{author}{\bibfnamefont{A.~D.} \bibnamefont{Speliotopoulos}},
  \bibinfo{journal}{Int. J. Mod. Phys. D} \textbf{\bibinfo{volume}{12}},
  \bibinfo{pages}{1627} (\bibinfo{year}{2003}).

\bibitem[{\citenamefont{Storey and Cohen-Tannoudji}(1994)}]{storey94}
\bibinfo{author}{\bibfnamefont{P.}~\bibnamefont{Storey}} \bibnamefont{and}
  \bibinfo{author}{\bibfnamefont{C.}~\bibnamefont{Cohen-Tannoudji}},
  \bibinfo{journal}{J. Phys. II France} \textbf{\bibinfo{volume}{4}},
  \bibinfo{pages}{1999} (\bibinfo{year}{1994}).

\end{thebibliography}

\end{document}